\documentclass{article}

 \usepackage[final,nonatbib]{neurips_2020}

\usepackage[utf8]{inputenc} 
\usepackage[T1]{fontenc}    
\usepackage{hyperref}       
\usepackage{url}            
\usepackage{booktabs}       
\usepackage{amsfonts}       
\usepackage{nicefrac}       
\usepackage{microtype}      
\usepackage{textgreek}

\usepackage{multirow}

\usepackage{bm}

\newcommand{\comment}[1]{}

\usepackage{graphicx}
\usepackage{subcaption} 
\usepackage[font=small,skip=0.3\baselineskip]{caption}
\usepackage{amsmath}
\usepackage{amssymb}
\usepackage{bm}
\usepackage[]{algorithm2e}

\usepackage{enumitem}

\usepackage[toc,page,header]{appendix}
\usepackage{minitoc}

\newcommand{\norm}[1]{\left\|#1\right\|}

\newcommand{\xb}{\mathbf{x}}

\newcommand{\wb}{\mathbf{w}}
\newcommand{\ub}{\mathbf{u}}
\newcommand{\vb}{\mathbf{v}}

\newcommand{\RR}{\mathbb{R}}

\newcommand{\cB}{\mathcal{B}}

\title{Practical application improvement to Quantum SVM:  theory to practice
}

\author{%
  Jae-Eun Park\textsuperscript{1}, Brian Quanz\textsuperscript{2}, Steve Wood\textsuperscript{2}, Heather Higgins\textsuperscript{1}, Ray Harishankar\textsuperscript{1} \\
  \textsuperscript{1}IBM Global Business Services - Quantum Consulting Center of Competence, \textsuperscript{2}IBM Research \\
  \texttt{\{parkje,blquanz,woodsp,hhiggs,harishan\}@us.ibm.com} \\
}

\begin{document}

\maketitle
\begin{abstract}

Quantum machine learning (QML) has emerged as an important area for Quantum applications, although useful QML applications would require many qubits. Therefore our paper is aimed at exploring the successful application of the Quantum Support Vector Machine (QSVM) algorithm while balancing several practical and technical considerations under the Noisy Intermediate-Scale Quantum (NISQ) assumption.  For the quantum SVM under NISQ, we use quantum feature maps to translate  data into quantum states and build the SVM kernel out of these quantum states, and further compare with classical SVM with radial basis function (RBF) kernels.   As data sets are more complex or abstracted in some sense, classical SVM with classical kernels leads to less accuracy compared to QSVM, as classical SVM with typical classical kernels cannot easily separate different class data. Similarly, QSVM should be able to provide competitive performance over a broader range of data sets including ``simpler'' data cases in which smoother decision boundaries are required to avoid any model variance issues (i.e., overfitting).  To bridge the gap between ``classical-looking'' decision boundaries and complex quantum decision boundaries, we propose to utilize general shallow unitary transformations to create feature maps with rotation factors to define a tunable quantum kernel, and added regularization to smooth the separating hyperplane model. We show in experiments that this allows QSVM to perform equally to SVM regardless of the complexity of the data sets and outperform in some commonly used reference data sets.

\end{abstract}

\section{Introduction}

Quantum computation is a computational paradigm based on the laws of quantum mechanics, which enables a huge leap forwards in processing power \cite{williams2010explorations}. By carefully exploiting quantum effects such as interference or entanglement, quantum computers are expected to efficiently solve particular hard problems that would be intractable for classical machines, with quantum advantages such as exponential speed-up \cite{shor1999polynomial,van2006quantum}. On the other hand,  Quantum Machine Learning (QML) brings in slightly different research elements from the intersection with classical Machine Learning (ML) while leveraging the computational advantage of quantum computation \cite{biamonte2017quantum,adcock2015advances}.  There are many aspects and algorithms of QML such as solving linear systems of equations, principle component analysis (QPCA), and support vector machines.  In this paper, we focus on the support vector machine (QSVM) model in particular.  Similar to support vector machines, the Quantum SVM (QSVM) algorithm applies to classification problems that require a feature-map implicitly specified by a kernel (i.e., a function representing the inner product in the mapped feature space).  Specifically, some prior work discusses cases in which kernel computation is not efficient classically as it would scale exponentially with the size of the problem (i.e., large number of features) \cite{havlivcek2019supervised,rebentrost2014quantum}.  Besides kernel computation speed-up, other potential advantages of QSVM could include improved analytical performance (e.g., improved model accuracy), model training speed-up, and privacy \cite{havlivcek2019supervised,rebentrost2014quantum}. Past work has shown the potential of exponential speed-up for training SVMs using quantum computing \cite{rebentrost2014quantum}, in particular going from polynomial complexity to logarithmic complexity using quantum techniques for inverting the kernel matrix.

Therefore, a remaining key challenge is to provide \emph{quantum kernel functions} - kernel functions that can be readily expressed and computed with quantum circuits, that can be useful for modeling with different types of data, as well as the practical application of QSVM to different datasets.
An important QSVM framework was introduced and implemented recently in IBM Qiskit \cite{havlivcek2019supervised}. The authors demonstrated a high dimensional feature map and kernel - readily computable with quantum circuits - hence we refer to it as a \emph{quantum kernel} for convenience. This kernel is extremely difficult for classical algorithms to duplicate, and classical kernels fail to achieve good performance with data complementary to the imposed feature space.  
However, although this quantum kernel represents a powerful feature map covering the complex vector (Hilbert) space, it can easily lead to overfitting and poor performance as well, especially for other datasets not well-suited to the imposed feature space including a spectrum of ``simpler'' datasets that classical kernels and SVM can perform well on.
Additionally, we are presently in the so-called "noisy intermediate-scale quantum" (NISQ) era due to the limitation of physical qubits and fidelity, which has ideated a hybrid approach for many algorithms such as Variational Quantum Eigensolver (VQE) \cite{peruzzo2014variational,wang2019accelerated}. As one approach,  QSVM can also adopt its hybrid approach as a kernel estimator and feature-map can be constructed using a quantum computer and the remaining steps will be done using a classical computer \cite{havlivcek2019supervised}.  This would provide a much more flexible framework for us to introduce some of the classical ML elements such as varying types of regularization, and we can implement QSVM easily for real applications.

Therefore, our paper is aimed at exploring the successful practical application of the QSVM algorithm balancing several practical and technical considerations under the NISQ assumption.  We mainly explore and discuss the analytical performance of QSVM over a spectrum of different data sets starting from simple data, in which classical SVM and kernels work well, but without appropriate application of unitary transformation, the quantum kernel shown in \cite{havlivcek2019supervised} could lead overfitting and poor performance. We vary this complexity to more complex datasets, up to the point of the complex, carefully constructed dataset suitable for quantum kernels in \cite{havlivcek2019supervised}.  To bridge the gap between ``classical looking'' decision boundaries and complex quantum state induced decision boundaries for QSVM, we utilize general shallow unitary transformations, introducing variations of the quantum kernel and added regularization to smooth the hyperplane classifiers. This allows QSVM to perform equally or better than SVM regardless of the complexity of data sets.  As a result, the classification performance of QSVM performs well over different data sets and in some case, it shows better performance.

 Our main contributions are: 
 
\noindent$\bullet$  We show the methodology to introduce general unitary transformations to construct various feature-maps with quantum circuits, extending the set of \emph{quantum kernels} / kernel configurations that can be applied for QSVM.  We further show this can benefit model performance, with different quantum kernel variations leading to better fit for different datasets.  As such the variation selection can be viewed as a tunable hyper-parameter (as is common with classical SVM kernels).  These new quantum kernel variations, along with introduced $\ell_2$ regularization factors to QSVM, allow the extended QSVM approach to perform equally good or better than classical SVM over different data sets, without the overfitting issue apparent in the original quantum kernel QSVM approach. 

\noindent$\bullet$  We demonstrate the possibility of QSVM outperforming classical SVM, while bridging the gap between the effectiveness of QSVM from the complex data domain to the simple Euclidian data domain, with practical application of QSVM to a spectrum of datasets.  We show the results of experiments comparing analytical performance (accuracy) of classical SVM with QSVM and the proposed modifications across data sets.  We also show how the quantum kernel variation that works best varies for different data sets, suggesting a possible relationship between data set types and optimal quantum kernel types.

\section{Methodology}
\label{sec:methodology}

\subsection{ Notation and Setting} A matrix is denoted by a bold capital letter, vector by bold lowercase letters, and sets with calligraphic font. Given a vector $\xb$, its $i$-th element is denoted by $x_i$. 
$\norm{\xb}_{\ell_p}$ is the $\ell_p$-norm of the vector $\xb$, defined as $(\sum_i x_i^p)^{1/p}$. 
For a set $\cB$, $|\cB|$ is regarded as the cardinality of this set. 
The inner product of vectors $\ub$ and $\vb$ is denoted by $\ub \cdot \vb$.

For simplicity in this work we focus on supervised binary classification, but the approach readily extends to other tasks like regression and multi-class classification, as with kernel-based learning in general \cite{scholkopf2002learning}.  For binary classification the task is to learn a function $f^* : \mathcal{X} \rightarrow \{ \pm 1 \}$ that correctly predicts the corresponding class label $y \in \{ \pm 1 \}$ given a corresponding input $\xb \in \mathcal{X}$, where samples $(\xb,y)$ from $\mathcal{X} \times
\mathcal {Y}$ follow some (unknown) probability distribution $Pr(\xb, y)$.   In the supervised setting, we have available a set of $n$ training samples $\mathcal{D}_s = \{ \{\xb_1,y_1\}, \ldots, \{\xb_n, y_n\} \}$ independently sampled from $Pr(\xb, y)$, and the goal is to learn $f^*$ such that it will generalize well to (accurately predict) subsequent samples from $Pr(\xb, y)$.

\subsection{Support Vector Machines and Kernel-based learning} 

One widely used, well-studied, and effective type of model for supervised classification is the Support Vector Machine (SVM) classifier \cite{cortes1995support,scholkopf2002learning}.
SVM classifiers approach supervised classification with a linear model, defining the prediction as a function of the inner product between $\xb $ and a weight vector $\wb \in \mathcal{X}$ : $f^*(\xb) = \operatorname{sign}(f( \xb) + b)$, where $f(\xb) = \wb \cdot \xb$.  This decision boundary corresponds to a particular hyperplane with orientation controlled by $\wb$ and offset controlled by $b$.  To enable good generalization, support vector machines aim to find a hyperplane that separates the data in different classes as much as possible (i.e., with maximum margin).  In order to account for inseparable data, a soft margin formulation is typically used in which a hinge loss is traded off with the norm of $\wb$ (which can also be seen as regularizing the model) when fitting the model.  
In order to enable approximating arbitrary nonlinear functions $f^*$ with this simple linear model a nonlinear mapping $\phi: \mathcal{X} \rightarrow \mathcal{Z}$ which maps $\xb \in \mathcal{X}$ to some potentially higher-dimensional space $\mathcal{Z}$ is used, and a hyperplane (i.e., linear decision boundary) is found in this transformed space.

Given training set $\mathcal{D}_s$, training the soft-margin SVM  amounts to solving the following optimization:
\begin{equation}
\label{eq:bin_svm_w}
\begin{array}{l ll}
 \operatorname{min.}  & \frac{1}{2}\norm{\wb}_{\ell_2} +C  \sum_{i=1}^{n}\epsilon_i \\
 {\rm s.t.} &  \epsilon_i \ge 0 , \text{and }  y_i(\wb \cdot \phi(\xb_i) + b) \ge 1 - \epsilon_i  \quad \forall i=1,...,n \\
\end{array}
 \end{equation}

Equation \ref{eq:bin_svm_w} can be equivalently expressed using the dual of the optimization problem \cite{scholkopf2002learning}, resulting in only inner products between transformed data points.  This allows the problem and solution to be expressed using only inner products of $\phi(\xb)$ vectors. The inner product in the transformed space $\mathcal{Z}$ can then be represented by a \emph{kernel function} defined in the input space, $K: \mathcal{X} \times
\mathcal {X} \rightarrow \RR$ such that $K(\ub,\vb) = \phi(\ub)^T \phi(\vb) $.  Therefore the problem can be expressed using only kernel functions in the input space without having to explicitly define the feature mapping $\phi()$ (which can be complex) or find $\wb$ in the possibly very high dimensional or arbitrary unknown space $\mathcal{Z}$.  This results in a functional formulation for the model that can be viewed as kernel basis functions weighted by $\bm{\beta}$, with upper constraint of $C$ acting as $\ell_1$ norm regularization on $\bm{\beta}$.

\begin{equation}
\label{eq:svm_dual}
\begin{array}{l ll}
 \operatorname{max.}  &  L(\bm{\beta}) = \sum_{i=1}^n \beta_i - \frac{1}{2}  \sum_{i=1}^n\sum_{j=1}^n \beta_i \beta_j y_i y_j K(\xb_i , \xb_j) \\
 {\rm s.t.} & 
 \sum_{i=1}^n \beta_i y_i = 0, \text{and } 0 \leq \beta_i \leq C
  \quad \forall i=1,...,n \\
\end{array}
 \end{equation}

\subsubsection{Importance of kernel selection and challenges of SVM}
Using the nonlinear, \emph{kernelized} version of SVM enables modeling arbitrarily complex and nonlinear functions, given an appropriate kernel choice.  However there are two key challenges.  First is the the polynomial scaling for training the model - the computation typically grows on the order of $n^3$ with number of data points $n$ (aside from specialized sub-optimal methods, e.g. \cite{tsang2005core}).  This has somewhat limited the applicability of and possibly the interest in SVM and kernel-based methods with growing data sizes.
Second, since the kernel inherently corresponds to the nonlinear transformation of the input data used to do the modeling, it is crucial to select the right kernel function for a given dataset \cite{scholkopf2002learning}.  There is a limited set of well-studied and commonly used classes of kernel functions (e.g., polynomial, and radial basis function kernels) that have a few hyper-parameters that also must be carefully tuned for the data.  For example radial basis function (RBF) kernels, with $K(\ub,\vb) =  \exp(-\norm{\ub - \vb}_{\ell_2}/h)$, are widely used but the kernel width $h$ must be carefully tuned to find the right fit to data.  The importance and difficulty of selecting an appropriate kernel has led to research on automatically learning a kernel function \cite{micchelli2005learning,gonen2011multiple,cortes2009learning} but this is also computationally challenging in general and often does not provide explicit kernels, or restricts the set of possible kernel functions in some way. 

\subsection{Quantum SVM and Quantum Kernels}
We argue that quantum computing has the potential to help address the challenges of SVM and kernel learning, if we can use quantum circuits to compute useful kernel functions and improve the scalability of fitting an SVM model.  Past work has shown the potential of exponential speedup for training SVMs using quantum computing \cite{rebentrost2014quantum}, in particular going from polynomial complexity to logarithmic complexity using quantum techniques for inverting the kernel matrix.

Therefore, the remaining key challenge is to provide \emph{quantum kernel functions} - kernel functions that can be readily expressed and computed with quantum circuits, that can be useful for modeling with different types of data. 
As the kernel function inherently reflects the transformed data space for modeling, different kernels and classes of kernels implicitly represent different types of transformations, and different kernels will be best suited for different data.

\subsubsection{Quantum kernels and decision boundaries}
\begin{figure}[ht]
	\centering
	\begin{subfigure}[t]{0.45\textwidth} 
		\includegraphics[width=\textwidth]{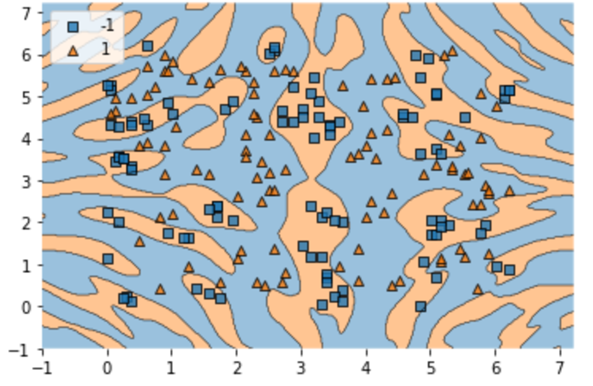}
		\caption{} 
		\label{fig:decision_boundary}
	\end{subfigure}
	\hspace{1em} 
	\begin{subfigure}[t]{0.3\textwidth} 
		\includegraphics[width=\textwidth]{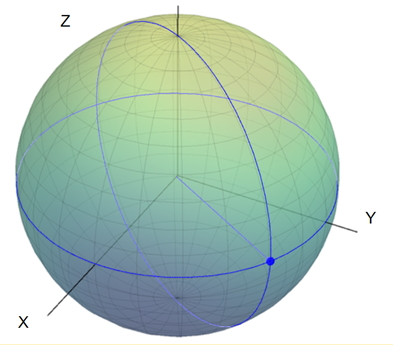}
		\caption{}
		\label{fig:bloch_sphere}
	\end{subfigure}
	\caption{{(a)} QSVM decision boundary. {(b)} The Bloch Sphere.}
	\label{fig:qsvm_and_bloch}
\end{figure}

Quantum SVM and kernels can exploit the higher dimensional space efficiently and can generate feature maps and subsequent decision boundaries that are difficult for classical kernel functions to match as shown in Figure \ref{fig:decision_boundary}, illustrating QSVM decision boundaries for a specialized data set designed to be separable using quantum kernels proposed in \cite{havlivcek2019supervised}. This can be understood by how the classical data are mapped and transformed into the Hilbert space using the Bloch Diagram in Figure \ref{fig:bloch_sphere}.  

However, we found this kernel as proposed in \cite{havlivcek2019supervised}, while effective for the complex designed data, does not yield accurate results on a spectrum of less complex and classical data sets (illustrated in subsequent sections).  We therefore propose an extension to the kernel, by introducing multiple rotations and rotational factors  ($\alpha$).
At the beginning to build the feature map, the classical features needs to be mapped into quantum data points, these data could be scaled and mapped into 0 to 2\textpi\space or be directly fed in the unitary transformation gate as shown in Equation \ref{eq:Feature Map}. 

\begin{equation}
\label{eq:Feature Map}
\begin{array}{l ll}
 
  U_{\phi(x)} = \exp(i\sum_{j=1}^n \alpha_j \phi_s(x) \prod \sigma_{j \in \{X,Y,Z\}})

\end{array}
\end{equation}

The $\alpha$ represent rotation factors, which control phase rotation depending on the feature values. The $\sigma$ represents unitary Pauli rotation transformation for X, Y and Z.  This reflects Ising like interactions (i.e. ZZ or YY) and non-interacting terms (i.e. Z, or Y).  These gates are preceded by the Hadamard gates applied to each qubit as in \cite{havlivcek2019supervised}.

\begin{figure}[ht]
	\centering
	\begin{subfigure}[t]{0.4\textwidth} 
		\includegraphics[width=\textwidth]{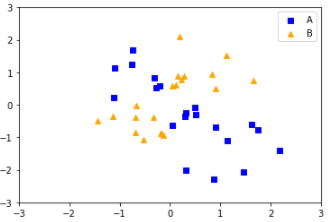}
		\caption{} 
	\end{subfigure}
	\hspace{1em} 
	\begin{subfigure}[t]{0.4\textwidth} 
		\includegraphics[width=\textwidth]{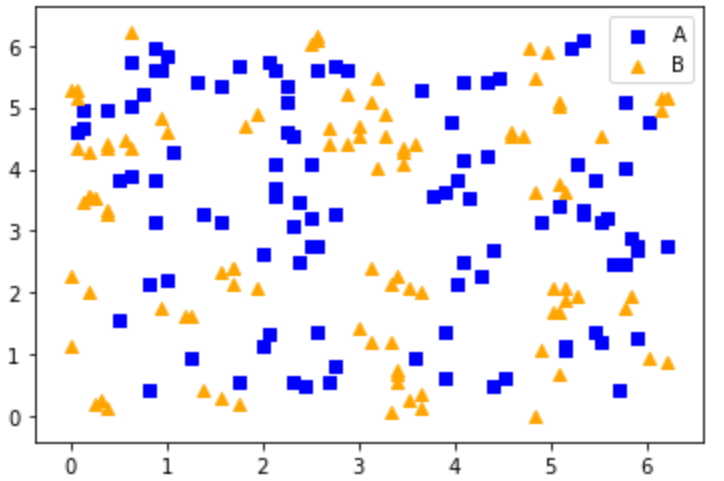}
		\caption{}
	\end{subfigure}
		\begin{subfigure}[t]{0.39\textwidth} 
		\includegraphics[width=\textwidth]{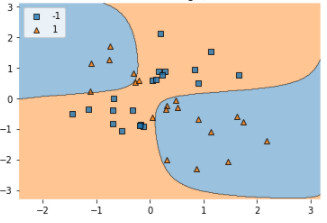}
		\caption{} 
	\end{subfigure}
	\hspace{1em} 
	\begin{subfigure}[t]{0.4\textwidth} 
		\includegraphics[width=\textwidth]{figures/F1_Ad_DB.png}
		\caption{}
	\end{subfigure}
	\caption{{(a)} XOR patterned data. {(b)}: Complex data. {(c)} QSVM decision boundary for XOR patterned data. {(d)}: QSVM decision boundary for complex data.}
	\label{fig:result1}
\end{figure}

Using different permutations of Pauli transformations and rotation factors $\alpha$, we can control how qubits revolve around the Bloch sphere presenting different complex amplitudes depending on input feature values and treat them as hyper-parameters to fit to given data.  This allows us to control the decision boundaries to be more representative of the data pattern without creating overly-complex decision boundaries. Also it allows the complex hyperplane to be created as needed based on the data set. To demonstrate the point, two artificial data sets were generated for the comparison of two extreme cases: randomly generated XOR-patterned data and another complex data \cite{havlivcek2019supervised}, and their corresponding decision boundaries are shown in Figure \ref{fig:result1} using our proposed quantum kernel. This demonstrates how flexible QSVM decision boundary can be.  Though two data sets have similar centroid locations for different labels, clearly different feature spaces and hyperplanes are required to separate and model these data labels successfully. Using the general unitary rotation in Equation \ref{eq:Feature Map}, we can create appropriate decision boundaries for these two distinct data sets by optimizing the rotation factors and combinations of Pauli gates depending on the complexity of the data sets. These gates in Equation \ref{eq:Feature Map} provide enough flexibility to match the complexity of complex decision boundaries but can also be simplified to be effective for simple data sets such as XOR.

\begin{figure}[ht]
	\centering
	\begin{subfigure}[t]{0.46\textwidth} 
		\includegraphics[width=\textwidth]{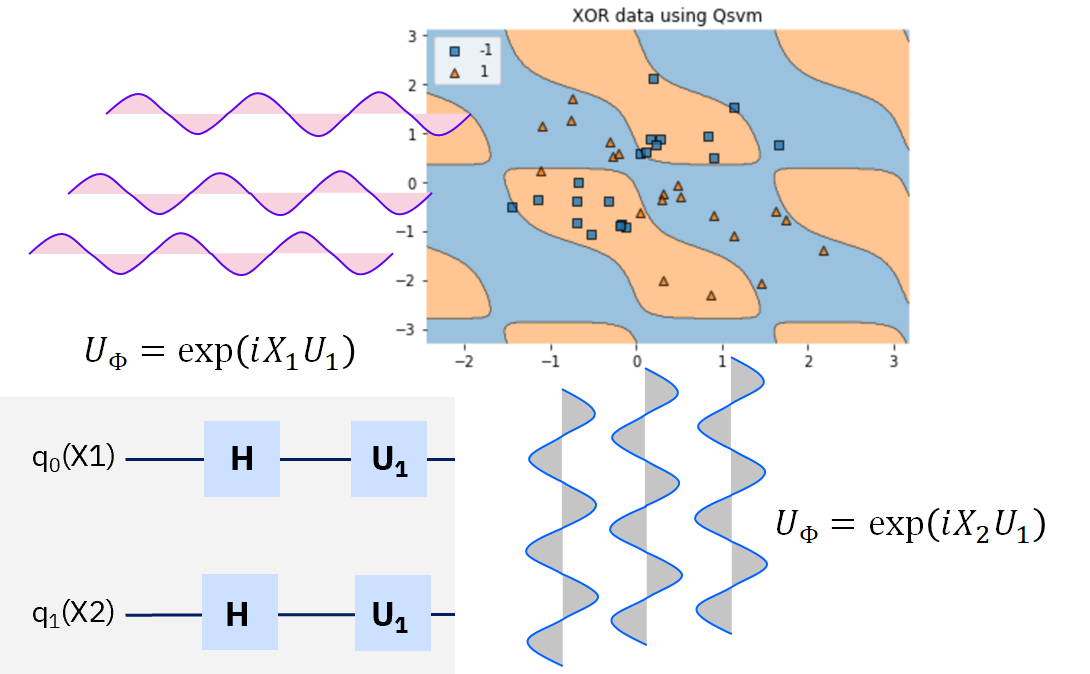}
		\caption{} 
		\label{fig:single_rotation}
	\end{subfigure}
	\hspace{1em} 
	\begin{subfigure}[t]{0.3\textwidth} 
		\includegraphics[width=\textwidth]{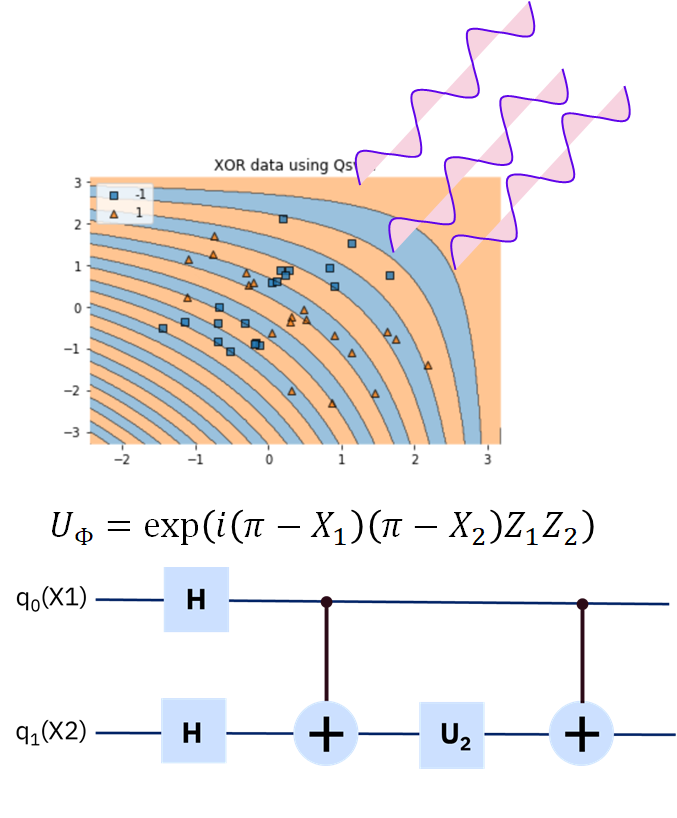}
		\caption{}
		\label{fig:entangled_rotation}
	\end{subfigure}

	\begin{subfigure}[t]{0.4\textwidth} 
		\includegraphics[width=\textwidth]{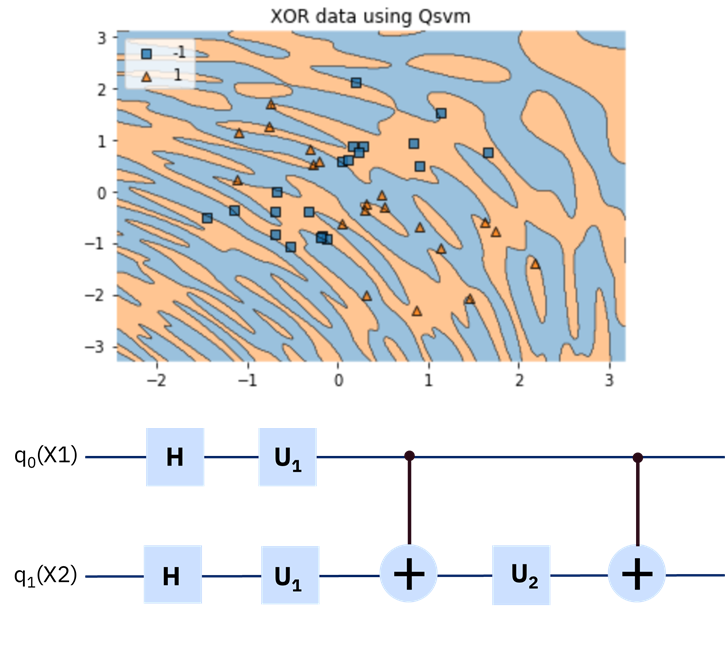}
		\caption{} 
		\label{fig:rotation_combination}
	\end{subfigure}
	\hspace{1em} 
	\caption{{(a)} Single unitary rotations by each factor. {(b)}: Entangled unitary rotations by two factors.{(c)}: Combination between (a) and (b).}
	\label{fig:rotation_illustration}
\end{figure}

   To give a high level intuition, single unitary rotation gates can provide a variety of feature maps that can be used to create simple boundary conditions.  This single unitary rotation will revolve along with the input data depending on how it is mapped and applied to the phase rotation of the gate as shown in Figure \ref{fig:single_rotation}.  If features need to interact, the interaction can be created through the entanglement (i.e. ZZ gate) and propagates diagonally through the two feature space (Figure \ref{fig:entangled_rotation}).  By combining these two single unitary and interacting transformation together, the complex and higher order of decision boundaries can be created shown in Figure \ref{fig:rotation_combination}.

\subsection{Regularization} 
Regularization is used to reduce generalization error and model complexity by including model complexity penalties in the training objective.  
Due to the hybrid approach for QSVM mentioned previously, we can more easily implement regularization as part of the optimization process as shown in Equation \ref{eq:new_loss} - defining new loss function $\hat{L}(\bm{\beta})$ to use in place of $L(\bm{\beta})$ in Equation \ref{eq:svm_dual}. Here we include the $\ell_1$ constraint as in the classical formulation (which can also be expressed as  $\ell_1$ norm penalty), as well as introducing an additional $\ell_2$ norm penalty. 

\begin{equation}
\label{eq:new_loss}
\begin{array}{l ll}
 

  \hat{L}(\bm{\beta})= L(\bm{\beta}) +\lambda_1 \norm{\bm{\beta}}_{\ell_1}+\lambda_2 \norm{\bm{\beta}}_{\ell_2}
  

\end{array}
\end{equation}

Given the flexibility of QSVM’s decision boundary, introducing proper regularization is an important factor to avoid over-fitting.  Introducing $\ell_2$ regularization can also help encourage smoother blending of the complex functionals, which may be particularly beneficial with these more complex kernel functions, vs. $\ell_1$ penalty which encourages sparse selection as opposed to smoothing.  Interestingly, we found the $\ell_2$ regularization  to have a much greater smoothing effect in our experiments.

\begin{figure}[ht]
	\centering
	\begin{subfigure}[t]{0.32\textwidth} 
		\includegraphics[width=\textwidth]{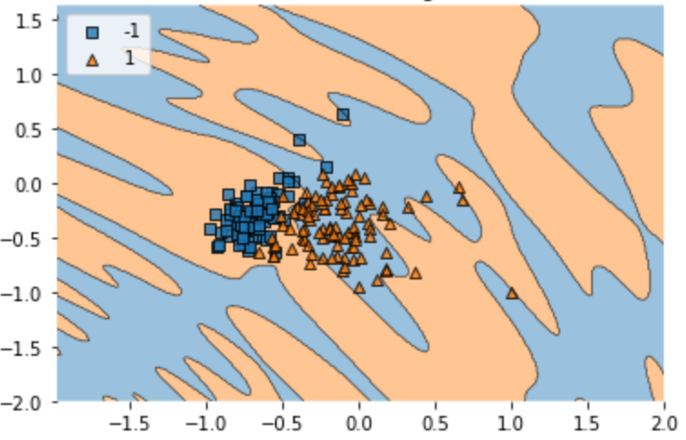}
		\caption{} 
		\label{fig:step1}
	\end{subfigure}\hfil 
	\begin{subfigure}[t]{0.34\textwidth} 
		\includegraphics[width=\textwidth]{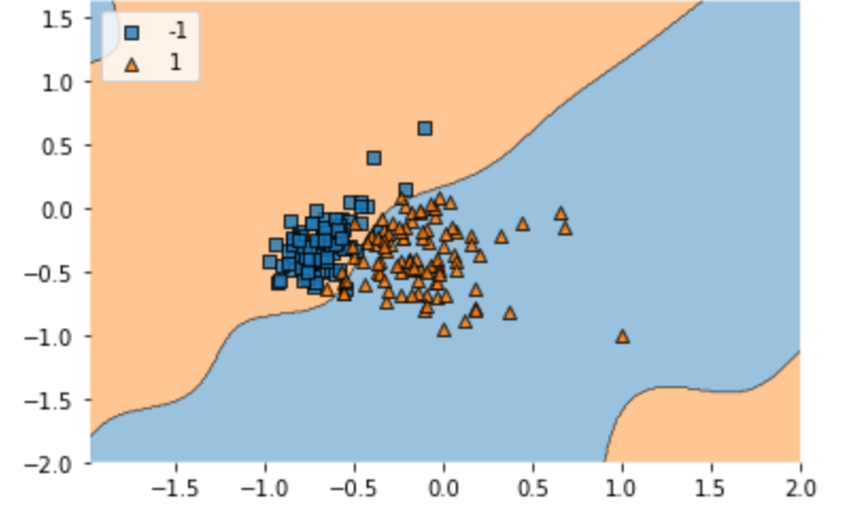}
		\caption{}
		\label{fig:step2}
	\end{subfigure}\hfil 
	\begin{subfigure}[t]{0.33\textwidth} 
		\includegraphics[width=\textwidth]{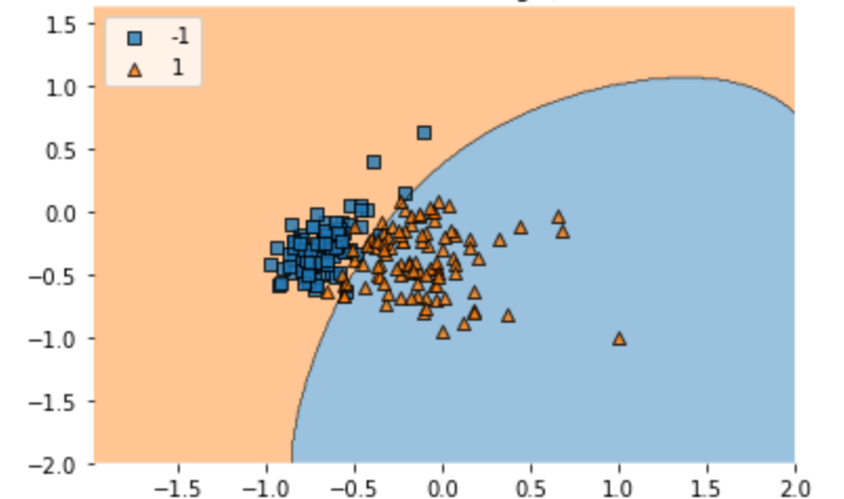}
		\caption{} 
		\label{fig:step3}
	\end{subfigure}
	
	\begin{subfigure}[t]{0.33\textwidth} 
		\includegraphics[width=\textwidth]{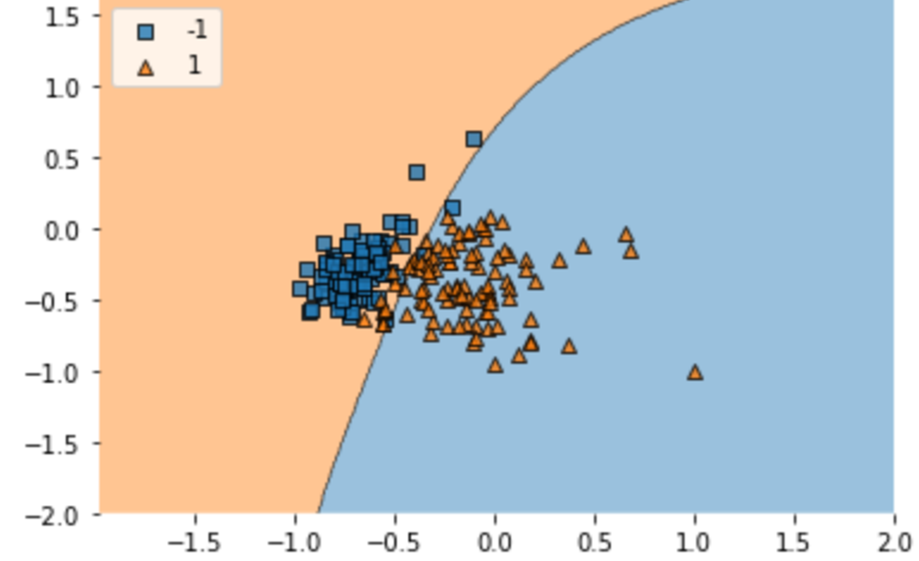}
		\caption{}
		\label{fig:step4}
	\end{subfigure}\hfil 
	\begin{subfigure}[t]{0.33\textwidth} 
		\includegraphics[width=\textwidth]{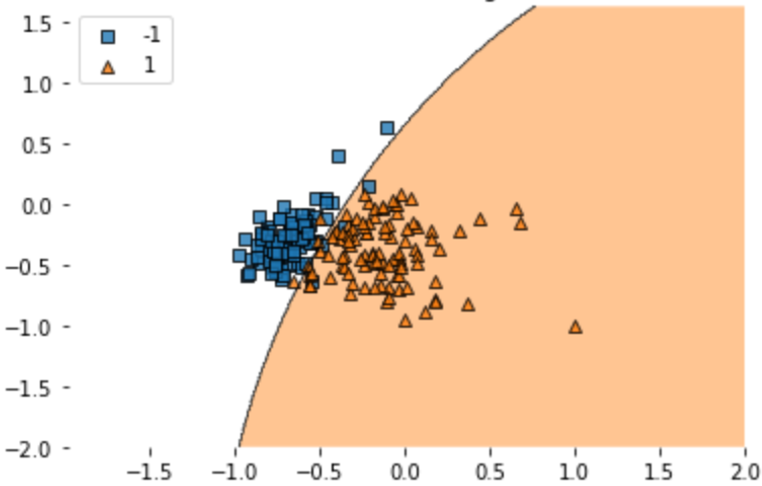}
		\caption{} 
		\label{fig:step5}
	\end{subfigure}
	
	\caption{ Breast cancer data from Python Sklearn datasets {(a)} Pauli Y YY with \textalpha=2. {(b)}: Pauli Y with \textalpha=2. {(c)} Pauli Y with \textalpha=1.{(d)} Pauli Y with \textalpha=1 and L2 regularization.{(e)}  Classical SVM.  }
	\label{fig:result2}
\end{figure}

From Figure \ref{fig:result2} we can see how the ``classical-looking'' (simpler) decision boundaries can be formed with the proposed Quantum kernel and SVM through appropriate gate implementation (Equation \ref{eq:Feature Map}) and  $\ell_2$ regularization factors (Equation \ref{eq:new_loss}) during the optimization step. The combination of Y and YY Pauli transformation can easily overfit this data set (Figure \ref{fig:step1}) and it can be gradually improved using a simple transformation and low rotation factor $\alpha$ (Figures \ref{fig:step2}-\ref{fig:step3}). Finally, we can see how regularization can improve the boundary and make the decision boundary of QSVM look like that of classical SVM (Figures \ref{fig:step4}-\ref{fig:step5}).  Here $\lambda_1 = 0$ and for Figure \ref{fig:step4} $\lambda_2$ was increased until there was a noticeable impact on the decision boundary.   Note, Figure \ref{fig:step1} illustrates how direct application of the type of quantum kernel approach from \cite{havlivcek2019supervised} may lead to overfitting on such simpler decision boundary data.  This along with Figures \ref{fig:step2}-\ref{fig:step4} suggests the need for the proposed modifications to enable quantum kernels to be effective for a wide variety of data.
Note here we loosely use terms like ``classical-looking'' to describe the intuitive idea of decision boundaries that are in some sense less complex (in terms of smoothness and minimal geometric separation of class groups) and more representative of boundaries typically derived from using classical kernels.  We leave precise definition and more investigation of how to characterize the difference in decision boundaries and specifics of what data works better with different quantum kernel variations to future work.

\section{Analytical performance}

As explained in Section \ref{sec:methodology}, the function of an SVM kernel is to implicitly transform data into a more suitable space for modeling. Different SVM algorithms use different types of kernel functions such as linear, nonlinear, polynominal, radial basis (RBF), and sigmoid, as well as special kernels for sequence data, graphs, text, and images.  Arguably the most commonly used type of kernel function is RBF due to its localization and finite response\cite{vert2004kernel}.  In this section, we present classification accuracy results for different types of quantum kernels for QSVM and compare to RBF for classical SVM as shown in Table \ref{table:svm_comp_results}.  Our motivation is to demonstrate the performance of QSVM could be comparable to classical SVM, in particular with the proposed extensions, and in certain cases, even better - i.e., as the data becomes too complex / no longer well-modeled with classical kernels.

For classical SVM and kernel results, we performed light hyper-parameter tuning on validation data to arrive at seemingly reasonably results close to best obtainable through tuning (e.g., varying hyper parameters and observing the impact on classification performance and decisions boundaries, and selecting those that showed best data fit).  More thorough tuning might result in somewhat improved scores, but the reported results should be representative of the type of performance we could obtain with the classical SVM kernel.  As mentioned we focused on the RBF kernel due to its wide use and general flexibility in modeling arbitrary and complex decision boundaries.  Further, initial experiments trying other common kernels did not yield improved results.

For this study, five different balanced data sets were prepared; two artificially generated data sets (XOR and previously described dataset of \cite{havlivcek2019supervised}) and three commonly used data sets (wine, breast cancer, and hand-written digits) from Scikit-Learn \cite{Scikit-Learn}, compressed into 2 dimensions using principle component analysis (PCA) as shown in Figure \ref{fig:datasets}, in order to create more complex looking data and decision boundaries. Because of the balanced data, we look at only the classification accuracy as a metric for the comparison, and held out 30\% of the data for testing.
XOR patterned data (\ref{fig:xor}) and the artificial complex data sets (\ref{fig:wine}-\ref{fig:digit}) were created as discussed. We can observe an interesting trend over different data sets; classical SVM performs gradually more poorly over the noticeably increasingly complex data sets such as compressed hand-written digits (\ref{fig:digit}) and the artificial complex data (\ref{fig:adhoc}).  However, the performance of QSVM can be consistent through different data sets with appropriate combination between unitary gate and rotation factor. In fact, we can see the better performance from QSVM over classical SVM in the compressed hand-written data set as well. This leads to an interesting point in that QSVM can potentially provide better performance if the underlying boundary is a complex one not captured well by traditional classical kernels, and further directly duplicating the quantum kernels could be intractable as mentioned previously \cite{havlivcek2019supervised}.  Therefore this argues for a potential analytical performance advantage from using QSVM as we propose, vs. classical SVM.

\newcommand{\mysize}{0.33}
\begin{figure}[ht]
	\centering
	\begin{subfigure}[t]{\mysize\textwidth} 
		\includegraphics[width=\textwidth]{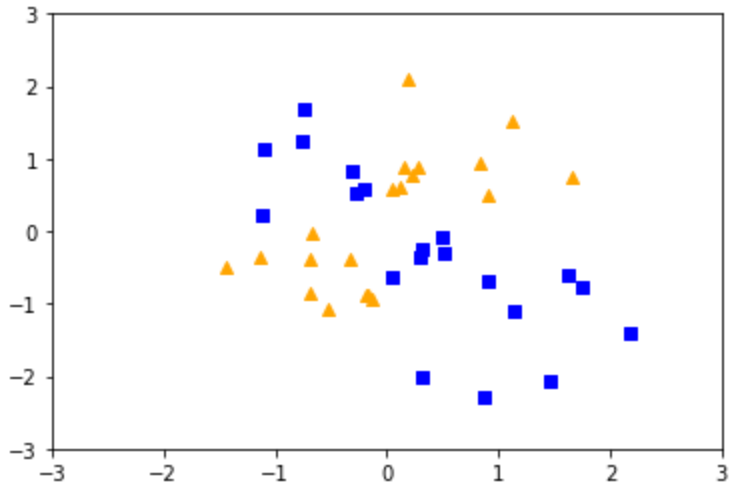}
		\caption{} 
		\label{fig:xor}
	\end{subfigure}\hfil 
	\begin{subfigure}[t]{\mysize\textwidth} 
		\includegraphics[width=\textwidth]{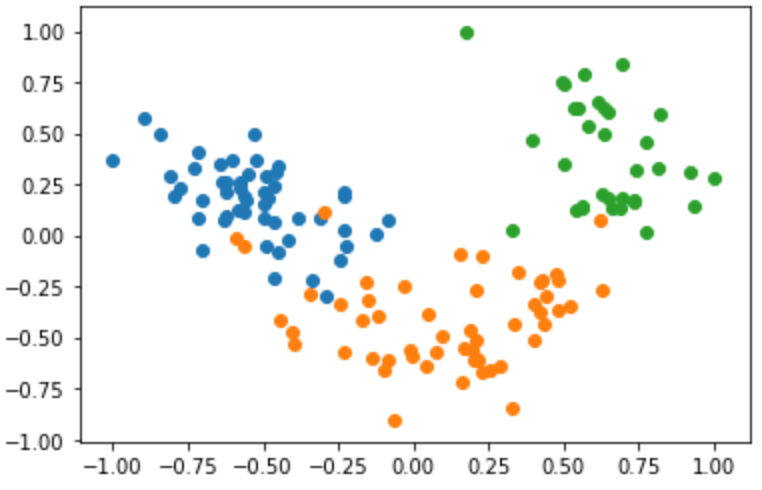}
		\caption{}
		\label{fig:wine}
	\end{subfigure}\hfil 
	\begin{subfigure}[t]{\mysize\textwidth} 
		\includegraphics[width=\textwidth]{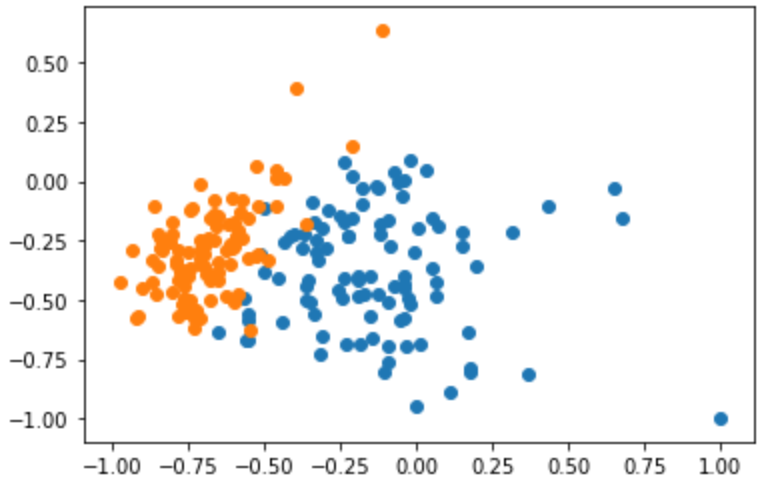}
		\caption{} 
		\label{fig:breast}
	\end{subfigure}

	\begin{subfigure}[t]{\mysize\textwidth} 
		\includegraphics[width=\textwidth]{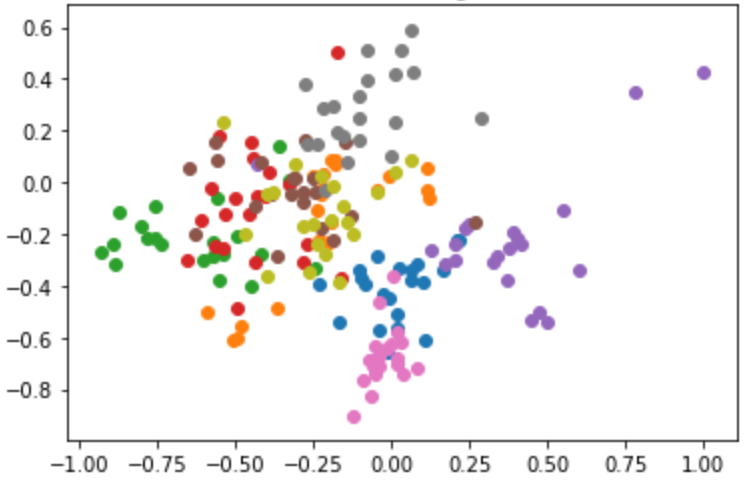}
		\caption{}
		\label{fig:digit}
	\end{subfigure}\hfil 
	\begin{subfigure}[t]{\mysize\textwidth} 
		\includegraphics[width=\textwidth]{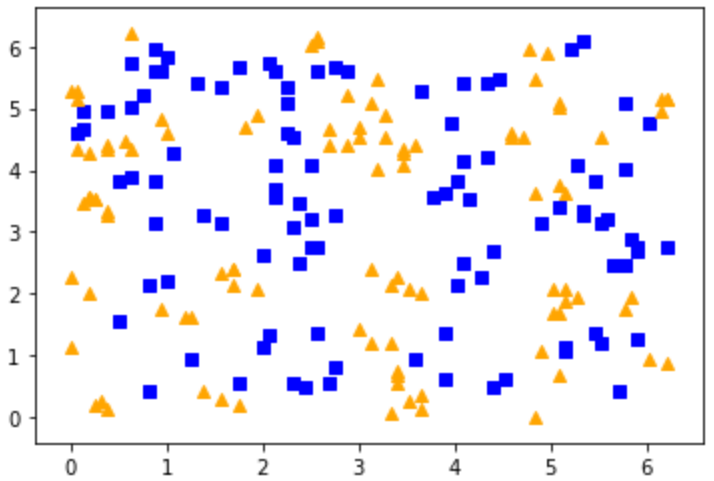}
		\caption{} 
		\label{fig:adhoc}
	\end{subfigure}
	\caption{{(a)} XOR patterned data. {(b)}: PCA compressed wine data. {(c)} PCA compressed breast cancer.{(d)} PCA compressed hand-written numbers.{(e)}  Artificial complex data.}
	\label{fig:datasets}
\end{figure}

\begin{table}[h!]
\centering
\caption{Comparison of SVM and Quantum SVM held-out test accuracy with different data sets. Only accuracy metrics are used as all data are balanced. The table is intended to show high level performance trends rather than comprehensive / exhausting benchmark comparison. }
\label{table:svm_comp_results}
\scriptsize
\begin{tabular}{  l | l | l | l | l | l | l  }
\multirow{2}{*}{Model} & \multirow{2}{*}{Feature Maps(function)} & 
\multicolumn{5}{ c}{Datasets} \\ 
& & XOR & Wines & Breast Cancer & Digits & Complex  \\ 
\hline \hline
Classical SVM 
 & RBF & $\textbf{99}$ & $95$ & $\textbf{96}$ & $43 $ & $5 $  \\ 
\hline
\multirow{5}{*}{Quantum SVM}
 & Pauli Y & $ \textbf{98}$ & $95$ & $\textbf{96}$& $43 $ & $66 $   \\
 & Pauli Z &  $ 80$ & $\textbf{97}$ & $95 $& $38 $ & $66 $  \\
 & Pauli YY & $ 67$ & $ 44 $ & $61 $& $20 $ & $52 $ \\
 & Pauli ZZ & $ 42$ & $ 67$ & $ 62$& $19 $ & $52 $ \\
 & Pauli Y YY & $ 82$ & $ \textbf{98}$ & $ 92$& $\textbf{48} $ & $88 $ \\
 & Pauli Z ZZ & $ 92$ & $97$ & $90 $& $\textbf{49} $ & $\textbf{100 }$  \\
 \hline

\end{tabular}
\end{table}


\section{Additional Related Work}

There is some related work exploring different aspects / approaches for QSVM or QML.  
A QSVM system based on an optimized HHL (Harrow, Hassidim, and Lloyd), quantum circuit with reduced circuit depth was presented to classify two-dimensional datasets that are linearly separable\cite{yang2019support}.
Two approaches for building a quantum model for QSVM were discussed using 2-dimensional mini-benchmark datasets.  The quantum device estimates inner products of quantum states to compute a classically intractable kernel. This kernel can be fed into any classical kernel method such as SVM. Also, a variational quantum circuit as a linear model was used to classify data explicitly in a Hilbert space \cite{schuld2019quantum}. 
From tensor networks for machine learning in the classical context, quantum computing approaches to both discriminative and generative learning, with circuits based on tree and matrix product state tensor networks, were proposed to show benefits with NISQ devices \cite{huggins2019towards}.

\section{Conclusion}
 
 We have discussed and demonstrated how to create decision boundaries using Quantum Support Vector Machines, which exploits a rich feature space. Due to the powerful feature transformation, finding appropriate kernel functions, which can be readily expressed and computed with simple quantum circuits, is important for analytical performance over various different data sets. To avoid overfitting and increase the applicability of the quantum kernel, we proposed quantum kernels using general unitary transformation with rotation factors along with added regularization for QSVM. This combination allows QSVM to perform consistently over different types of data sets.  In certain data sets when data are complex / mixed up such that the classical kernel approaches might fail, QSVM can perform better.
 
 For simplicity in this work experiments were performed on low-dimensional data.  One key area of future work is to apply this approach to higher-dimensional data as well, i.e., data having larger numbers of features.  We plan to perform similar experiments on more real-world data.  One challenge with this is that for larger number of features, larger quantum circuits and number of qubits could be needed, and even simulation of these larger systems with classical computing can become more compute-intensive.  This challenge will be mitigated as effective number of qubits in quantum systems increases in the future.  Another area of future work is determining and evaluating speed-ups from quantum vs. classical approaches.

\bibliographystyle{IEEEtran}
{\small
\bibliography{reference}}

\begin{thebibliography}{10}
\providecommand{\url}[1]{#1}
\csname url@samestyle\endcsname
\providecommand{\newblock}{\relax}
\providecommand{\bibinfo}[2]{#2}
\providecommand{\BIBentrySTDinterwordspacing}{\spaceskip=0pt\relax}
\providecommand{\BIBentryALTinterwordstretchfactor}{4}
\providecommand{\BIBentryALTinterwordspacing}{\spaceskip=\fontdimen2\font plus
\BIBentryALTinterwordstretchfactor\fontdimen3\font minus
  \fontdimen4\font\relax}
\providecommand{\BIBforeignlanguage}[2]{{%
\expandafter\ifx\csname l@#1\endcsname\relax
\typeout{** WARNING: IEEEtran.bst: No hyphenation pattern has been}%
\typeout{** loaded for the language `#1'. Using the pattern for}%
\typeout{** the default language instead.}%
\else
\language=\csname l@#1\endcsname
\fi
#2}}
\providecommand{\BIBdecl}{\relax}
\BIBdecl

\bibitem{williams2010explorations}
C.~P. Williams, \emph{Explorations in quantum computing}.\hskip 1em plus 0.5em
  minus 0.4em\relax Springer Science \& Business Media, 2010.

\bibitem{shor1999polynomial}
P.~W. Shor, ``Polynomial-time algorithms for prime factorization and discrete
  logarithms on a quantum computer,'' \emph{SIAM review}, vol.~41, no.~2, pp.
  303--332, 1999.

\bibitem{van2006quantum}
W.~Van~Dam, S.~Hallgren, and L.~Ip, ``Quantum algorithms for some hidden shift
  problems,'' \emph{SIAM Journal on Computing}, vol.~36, no.~3, pp. 763--778,
  2006.

\bibitem{biamonte2017quantum}
J.~Biamonte, P.~Wittek, N.~Pancotti, P.~Rebentrost, N.~Wiebe, and S.~Lloyd,
  ``Quantum machine learning,'' \emph{Nature}, vol. 549, no. 7671, pp.
  195--202, 2017.

\bibitem{adcock2015advances}
J.~Adcock, E.~Allen, M.~Day, S.~Frick, J.~Hinchliff, M.~Johnson,
  S.~Morley-Short, S.~Pallister, A.~Price, and S.~Stanisic, ``Advances in
  quantum machine learning,'' \emph{arXiv preprint arXiv:1512.02900}, 2015.

\bibitem{havlivcek2019supervised}
V.~Havl{\'\i}{\v{c}}ek, A.~D. C{\'o}rcoles, K.~Temme, A.~W. Harrow, A.~Kandala,
  J.~M. Chow, and J.~M. Gambetta, ``Supervised learning with quantum-enhanced
  feature spaces,'' \emph{Nature}, vol. 567, no. 7747, pp. 209--212, 2019.

\bibitem{rebentrost2014quantum}
P.~Rebentrost, M.~Mohseni, and S.~Lloyd, ``Quantum support vector machine for
  big data classification,'' \emph{Physical review letters}, vol. 113, no.~13,
  p. 130503, 2014.

\bibitem{peruzzo2014variational}
A.~Peruzzo, J.~McClean, P.~Shadbolt, M.-H. Yung, X.-Q. Zhou, P.~J. Love,
  A.~Aspuru-Guzik, and J.~L. O’brien, ``A variational eigenvalue solver on a
  photonic quantum processor,'' \emph{Nature communications}, vol.~5, p. 4213,
  2014.

\bibitem{wang2019accelerated}
D.~Wang, O.~Higgott, and S.~Brierley, ``Accelerated variational quantum
  eigensolver,'' \emph{Physical review letters}, vol. 122, no.~14, p. 140504,
  2019.

\bibitem{scholkopf2002learning}
B.~Sch{\"o}lkopf, A.~J. Smola, F.~Bach \emph{et~al.}, \emph{Learning with
  kernels: support vector machines, regularization, optimization, and
  beyond}.\hskip 1em plus 0.5em minus 0.4em\relax MIT press, 2002.

\bibitem{cortes1995support}
C.~Cortes and V.~Vapnik, ``Support-vector networks,'' \emph{Machine learning},
  vol.~20, no.~3, pp. 273--297, 1995.

\bibitem{tsang2005core}
I.~W. Tsang, J.~T. Kwok, and P.-M. Cheung, ``Core vector machines: Fast svm
  training on very large data sets,'' \emph{Journal of Machine Learning
  Research}, vol.~6, no. Apr, pp. 363--392, 2005.

\bibitem{micchelli2005learning}
C.~A. Micchelli and M.~Pontil, ``Learning the kernel function via
  regularization,'' \emph{Journal of machine learning research}, vol.~6, no.
  Jul, pp. 1099--1125, 2005.

\bibitem{gonen2011multiple}
M.~G{\"o}nen and E.~Alpayd{\i}n, ``Multiple kernel learning algorithms,''
  \emph{The Journal of Machine Learning Research}, vol.~12, pp. 2211--2268,
  2011.

\bibitem{cortes2009learning}
C.~Cortes, M.~Mohri, and A.~Rostamizadeh, ``Learning non-linear combinations of
  kernels,'' in \emph{Advances in neural information processing systems}, 2009,
  pp. 396--404.

\bibitem{vert2004kernel}
J.-P. Vert, K.~Tsuda, and B.~Sch{\"o}lkopf, ``Kernel methods in computational
  biology, chapter a primer on kernel methods,'' 2004.

\bibitem{Scikit-Learn}
F.~Pedregosa, G.~Varoquaux, A.~Gramfort, V.~Michel, B.~Thirion, O.~Grisel,
  M.~Blondel, P.~Prettenhofer, R.~Weiss, V.~Dubourg, J.~Vanderplas, A.~Passos,
  D.~Cournapeau, M.~Brucher, M.~Perrot, and E.~Duchesnay, ``Scikit-learn:
  Machine learning in {P}ython,'' \emph{Journal of Machine Learning Research},
  vol.~12, pp. 2825--2830, 2011.

\bibitem{yang2019support}
J.~Yang, A.~J. Awan, and G.~Vall-Llosera, ``Support vector machines on noisy
  intermediate scale quantum computers,'' \emph{arXiv preprint
  arXiv:1909.11988}, 2019.

\bibitem{schuld2019quantum}
M.~Schuld and N.~Killoran, ``Quantum machine learning in feature hilbert
  spaces,'' \emph{Physical review letters}, vol. 122, no.~4, p. 040504, 2019.

\bibitem{huggins2019towards}
W.~Huggins, P.~Patil, B.~Mitchell, K.~B. Whaley, and E.~M. Stoudenmire,
  ``Towards quantum machine learning with tensor networks,'' \emph{Quantum
  Science and technology}, vol.~4, no.~2, p. 024001, 2019.

\end{thebibliography}

\clearpage

\end{document}